\documentclass{ifacconf}

\usepackage{graphicx}      % include this line if your document contains figures
\usepackage{natbib}        % required for bibliography
\usepackage{amsmath}
\usepackage{amssymb}
\usepackage{mathrsfs}
\usepackage{lipsum} 
\usepackage{xcolor}

\usepackage{tikz}
\usetikzlibrary{arrows}
\usetikzlibrary{calc}
\usetikzlibrary{matrix}
\usetikzlibrary{fit}
\usetikzlibrary{shapes.arrows}

\newcommand{\tens}[1]{\boldsymbol{\mathcal{#1}}}
\newcommand{\tenselem}[1]{\mathcal{#1}}

\newcommand{\matr}[1]{\boldsymbol{#1}}
\newcommand{\mA}{\matr{A}}
\newcommand{\mB}{\matr{B}}
\newcommand{\mC}{\matr{C}}
\newcommand{\mH}{\matr{H}}
\newcommand{\mI}{\matr{I}}
\newcommand{\mP}{\matr{P}}

\newcommand{\mZ}{\matr{Z}}

\newcommand{\bmx}{\begin{bmatrix}}
\newcommand{\emx}{\end{bmatrix}}
\newcommand{\bsm}{\left[\begin{smallmatrix}}
\newcommand{\esm}{\end{smallmatrix}\right]}

\newcommand{\vect}[1]{\boldsymbol{#1}}

\newcommand{\cpd}[3]{[\![#1, #2,#3]\!]}

\newcommand{\kr}{\odot}     % Khatri-Rao product
\newcommand{\kron}{\mathop{\boxtimes}}   % Kronecker product
\newcommand{\con}{\mathop{\bullet}}   % Contraction
\newcommand{\T}{{\sf T}}        % transposition
\makeatletter
\newcommand{\vecl}[1]{\mathop{\operator@font vec}\{#1\}}
\newcommand{\rank}[1]{\mathop{\operator@font rank}\{#1\}}
\newcommand{\colrank}[1]{\mathop{\operator@font colrank}\{#1\}}
\newcommand{\krank}[1]{\mathop{\operator@font krank}\{#1\}}
\newcommand{\trace}[1]{\mathop{\operator@font trace}\{#1\}}
\newcommand{\Diag}[1]{\mathop{\operator@font Diag}\{#1\}}    % a diagonal matrix
\newcommand{\diag}[1]{\mathop{\operator@font diag}\{#1\}}    % a vector
\newcommand{\Span}[1]{\mathop{\operator@font Span}\{#1\}}    % a space
\newcommand{\argmin}{\mathop{\operator@font argmin}}

\newcommand{\cond}[1]{\mathop{\operator@font cond}\{#1\}}

\makeatother
\newcommand{\pt}{\mu}

\newcommand{\RR}{\mathbb{R}}
\newcommand{\CC}{\mathbb{C}}

\newcommand{\contr}[1]{\con_{#1}}

\newcommand{\ProjHD}[2]{\mathscr{P}_{#1,#2}}

\definecolor{darkgreen}{rgb}{0,0.6,0}
\definecolor{darkred}{rgb}{0.6,0,0}

\begin{document}

% Define lengths
\newlength{\wLTI}
\newlength{\wSNL}
\newlength{\wBigSNL}
\newlength{\wVeryBigSNL}
\newlength{\wModel}
\newlength{\wNamedSignal}
\newlength{\wShortNamedSignal}
\newlength{\wUnnamedSignal}
\newlength{\hLTI}
\newlength{\dyLTI}
\newlength{\sSum}
\newlength{\dySystemModel}
\newlength{\dySystemBLA}
\newlength{\marginBoundingBox}
\newlength{\wDelimiter}
\newlength{\dxArrowMatrix}
\newlength{\wArrow}
\newlength{\sCube}
\newlength{\dxFactorMatrix}

% Set lengths
\setlength{\wLTI}{3em}
\setlength{\wSNL}{\wLTI}
\setlength{\wBigSNL}{4.5em}
\setlength{\wVeryBigSNL}{5.5em}
\setlength{\wModel}{\wLTI}
\setlength{\wNamedSignal}{3em}
\setlength{\wShortNamedSignal}{2.5em}
\setlength{\wUnnamedSignal}{1em}
\setlength{\hLTI}{1.5em}
\setlength{\dyLTI}{1em}
\setlength{\sSum}{1em}
\setlength{\dySystemModel}{2\dyLTI}
\setlength{\dySystemBLA}{1.5\dyLTI}
\setlength{\marginBoundingBox}{0.5ex}
\setlength{\wDelimiter}{0.25em}
\setlength{\dxArrowMatrix}{0.5em}
\setlength{\wArrow}{0.75em}
\setlength{\sCube}{2em}
\setlength{\dxFactorMatrix}{1.5em}

% Define colors
\colorlet{LTIcolor}{red}
\colorlet{SNLcolor}{blue}
\colorlet{MODELcolor}{black}
\colorlet{RMScolor}{green!70!black}
\colorlet{PARAMcolor}{red}
\colorlet{ANNOTcolor}{blue}
\colorlet{HELPcolor}{gray!50}
\colorlet{MEANcolor}{black}
\colorlet{WORSTcolor}{red}
\colorlet{BESTcolor}{green!70!black}
\colorlet{OVERMODELINGcolor}{blue}
\colorlet{UNDERMODELING1color}{green!70!black}
\colorlet{UNDERMODELING2color}{red}
\colorlet{STEPcolor}{blue}
\colorlet{SLICEcolor}{gray!40}
\colorlet{SLICE2color}{white}

% Define constants
\def \ScaleFactor{5}

% Define block styles
\tikzstyle{SISO_LTI} 	= [draw, fill=none, text centered, minimum height=\hLTI, minimum width=\wLTI, LTIcolor]
\tikzstyle{MIMO_SNL} = [draw, fill=none, text centered, minimum height=(3\hLTI+2\dyLTI), minimum width=\wBigSNL, 
														SNLcolor]
\tikzstyle{SISO_SNL}	= [draw, fill=none, text centered, minimum height=\hLTI, minimum width=\wSNL, SNLcolor]
\tikzstyle{model}	= [draw, fill=none, text centered, minimum height=\hLTI, minimum width=\wModel, MODELcolor]
\tikzstyle{sum} 	= [draw, fill=none, text centered, minimum size=\sSum, circle]

\begin{frontmatter}

\title{Low-rank tensor recovery\\ for Jacobian-based Volterra identification  \\ 
of parallel Wiener-Hammerstein systems} 
% Title, preferably not more than 10 words.

%\thanks[footnoteinfo]{ \fin }

\author[First]{Konstantin Usevich} 
\author[Second]{Philippe Dreesen}
\author[Third]{Mariya Ishteva} 

\address[First]{Universit\'e de Lorraine, 
   CNRS, CRAN,  Nancy, France (e-mail: konstantin.usevich@univ-lorraine.fr).}
\address[Second]{KU Leuven, Dept.~Electrical Engineering (ESAT),
STADIUS Center for Dynamical Systems, Signal Processing and Data
Analytics, Belgium\\ (e-mail: philippe.dreesen@gmail.com)}
\address[Third]{KU Leuven, Department of Computer Science, ADVISE-NUMA, campus Geel, Belgium (e-mail: mariya.ishteva@kuleuven.be)}

\begin{abstract}                % Abstract of not more than 250 words.
We consider the problem of identifying a parallel Wiener-Hammerstein structure from Volterra kernels. Methods based on Volterra kernels typically resort to coupled tensor decompositions of the kernels. However, in the case of parallel Wiener-Hammerstein  systems, such methods require nontrivial constraints on the factors of the decompositions. In this paper, we propose an entirely different approach: by using special sampling (operating) points for the Jacobian of the nonlinear map from past inputs to the output, we can show that the Jacobian matrix becomes a linear projection of a  tensor whose rank is equal to the number of branches. This representation allows us to solve the identification problem as a tensor recovery problem.
\end{abstract}


\end{frontmatter}
%===============================================================================

\section{Introduction}
Nonlinear identification methods that go beyond  the well-established linear system identification tools \citep{pintelon2012,ljung1999,katayama2005}, are steadily gaining research attention in recent years.
Advances in nonlinear modeling tools, combined with the ever increasing computing power allows for the exploration of nonlinear models that account for nonlinear effects that occur when pushing systems outside of their linear operating regions. 
There is a host of procedures that range from simple extensions of linear models, over nonlinear state space modeling (possibly using particle filtering), to variations on neural network architectures, each of which typically require tailored nonconvex optimization methods. 
While such models may provide satisfactory prediction results, their internal workings are often hard to assess, which  makes them  difficult to use and interpret. 

The current paper considers a combination of two promising nonlinear models (block-oriented  models and Volterra series), and  aims at combining their advantages  while avoiding the drawbacks. 
Block-oriented systems are composed as interconnections of linear time-invariant system blocks and static nonlinearities such as the well-known Wiener, Hammerstein, Wiener-Hammerstein and Hammerstein-Wiener systems \citep{giri2010}. 
A block-oriented system description strikes a balance between flexibility and model interpretability: the model accounts for (strong) nonlinearities in its description, but stays close to the familiar linear world and allows for a transparent understanding of its workings. 
Nevertheless, block-oriented system identification methods rely on heuristics and nonconvex optimization routines \citep{schoukens2017} to find the parameters, which may cause difficulties.
Volterra series models, on the other hand, can be viewed as nonlinear extensions of the well-known convolution operation of the input signal with the (finite) impulse response: in the Volterra description, the output is defined as a polynomial function of (delayed) inputs (as opposed to the output being a linear function of delayed inputs in the case of linear systems). 
A major advantage is that the Volterra model is linear in the parameters and its identification can be posed as a least-squares problem \citep{birpoutsoukis2017}. 
Unfortunately, due to the polynomial structure, the number of coefficients grows very quickly as the polynomial degree increases. In addition, the model does not allow for an intuitive understanding of its internal operation. 

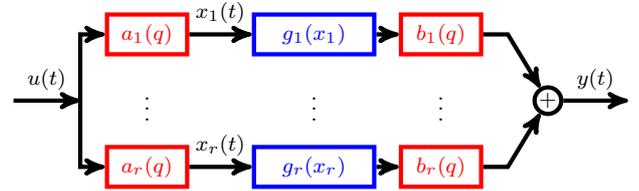
\begin{figure}
\begin{center}
\begin{tikzpicture}[font=\small, >=stealth', ultra thick, auto]
\useasboundingbox (-0.5,-1.5) rectangle (9.5,1.5);
	% general coordinates
	\node (u_start)	[coordinate]{};
	\node (u_end)		[coordinate, right of=u_start, node distance=\wShortNamedSignal]{};
	\node (u1_start)	[coordinate, above of=u_end, node distance=(\hLTI+\dyLTI)]{};
	\node (u3_start)	[coordinate, below of=u_end, node distance=(\hLTI+\dyLTI)]{};
	\node (front_LTI1_c)	[coordinate, right of=u1_start, node distance=(\wUnnamedSignal+\wLTI/2)]{};
	\node (front_LTI2_c)	[coordinate, right of=u_end, node distance=(\wUnnamedSignal+\wLTI/2)]{};
	\node (front_LTI3_c)	[coordinate, right of=u3_start, node distance=(\wUnnamedSignal+\wLTI/2)]{};
	\node (SNL1_c)	[coordinate, right of=front_LTI1_c, node distance=(\wLTI/2+\wShortNamedSignal+\wBigSNL/2)]{};
	\node (SNL2_c)	[coordinate, right of=front_LTI2_c, node distance=(\wLTI/2+\wShortNamedSignal+\wBigSNL/2)]{};
	\node (SNL3_c)	[coordinate, right of=front_LTI3_c, node distance=(\wLTI/2+\wShortNamedSignal+\wBigSNL/2)]{};
	\node (back_LTI1_c)	[coordinate, right of=SNL1_c, node distance=(\wBigSNL/2+\wUnnamedSignal+\wLTI/2)]{};
	\node (back_LTI2_c)	[coordinate, right of=SNL2_c, node distance=(\wBigSNL/2+\wUnnamedSignal+\wLTI/2)]{};
	\node (back_LTI3_c)	[coordinate, right of=SNL3_c, node distance=(\wBigSNL/2+\wUnnamedSignal+\wLTI/2)]{};
	\node (y1_end)	[coordinate, right of=back_LTI1_c, node distance=(\wLTI/2+\wUnnamedSignal)]{};
	\node (y2_end)	[coordinate, right of=back_LTI2_c, node distance=(\wLTI/2+\wUnnamedSignal)]{};
	\node (y3_end)	[coordinate, right of=back_LTI3_c, node distance=(\wLTI/2+\wUnnamedSignal)]{};
	\node (sum_c)		[coordinate, right of=y2_end, node distance=(\wUnnamedSignal+\sSum/2)]{};
	\node (y_end)		[coordinate, right of=sum_c, node distance=(\sSum/2+\wShortNamedSignal)]{};

	% specific nodes
	\node (front_LTI1)	at (front_LTI1_c) 	[SISO_LTI]{$a_1(q)$};
	\node (front_LTI2)	at (front_LTI2_c) 	[SISO_LTI, draw=none, black]{\vdots};
	\node (front_LTI3)	at (front_LTI3_c) 	[SISO_LTI]{$a_r(q)$};
	\node (SNL1)		at (SNL1_c)		[SISO_SNL, minimum width=\wBigSNL]{};
	\node 			at (SNL1)		[SNLcolor]{$g_1(x_1)$};
	\node (SNL2)		at (SNL2_c)		[SISO_SNL, minimum width=\wBigSNL, draw=none, black]{\vdots};
	\node (SNL3)		at (SNL3_c)		[SISO_SNL, minimum width=\wBigSNL]{};
	\node 			at (SNL3)		[SNLcolor]{$g_r(x_r)$};
	\node (back_LTI1)	at (back_LTI1_c) 	[SISO_LTI]{$b_1(q)$};
	\node (back_LTI2)	at (back_LTI2_c) 	[SISO_LTI, draw=none, black]{\vdots};
	\node (back_LTI3)	at (back_LTI3_c) 	[SISO_LTI]{$b_r(q)$};
	\node (sum)		at (sum_c)		[sum]{};
	\node 			at (sum)		{$+$};
		
	% signals
	\path[-]	(u_start)	-- node {$u(t)$}		(u_end);
	\path[-]	(sum)		-- node {$y(t)$}		(y_end);
	\path[-] 	(front_LTI1)	-- node {${x}_1(t)$}	(SNL1);
	\path[-] 	(front_LTI3)	-- node {${x}_r(t)$}	(SNL3);

	% arrows
	\draw[->] 	(u_start)	-- 	(u_end);
	\draw[<->] 	(front_LTI1)	-- 	(u1_start)	-- 	(u3_start)	-- 	(front_LTI3);
	\draw[->] 	(front_LTI1)	-- 	(SNL1);
	\draw[->] 	(front_LTI3)	-- 	(SNL3);
	\draw[->] 	(SNL1)		-- 	(back_LTI1);
	\draw[->] 	(SNL3)		-- 	(back_LTI3);
	\draw[->] 	(back_LTI1)	-- 	(y1_end)	-- 	(sum);
	\draw[->] 	(back_LTI3)	-- 	(y3_end)	-- 	(sum);
	\draw[->] 	(sum)		-- 	(y_end);
\end{tikzpicture}
\caption{Parallel Wiener-Hammerstein system.} 
\label{fig:pwh}
\end{center}
\end{figure}

In this article, we are interested in identification of discrete-time parallel Wiener-Hammerstein systems, see Fig.~\ref{fig:pwh}.
Each branch of  such a system has a Wiener-Hammerstein structure, i.e., a static nonlinearity sandwiched in between two linear time-invariant (LTI)  blocks.
Parallel Wiener-Hammerstein models have improved approximation properties  as opposed to single branch Wiener-Hammerstein models \citep{palm1979}. 
However, identification of a parallel Wiener-Hammerstein structure is particularly challenging, see \citet{schoukens2017}.
For example, the frequency-domain methods  \citep{schoukens2017} suffer from the pole/zero mixing of the Wiener and Hammerstein filters, and thus require a computationally heavy  pole/zero splitting procedure.

The method that we present in this article starts from estimating Volterra kernels, which can be readily viewed as higher-order symmetric tensors containing the polynomial coefficients. 
Existing methods that aim at finding block-oriented models from the Volterra kernels resort to coupled tensor decompositions of  Volterra kernels \citep{favier-structuredmatrix} and require nontrivial constraints on the factors of the tensor decomposition for parallel Wiener-Hammerstein case \citep{DreWesEtAl2017,WesIshDreEtAl17,dreesen2021sysid}. 
In this paper, we propose an entirely different approach: by choosing special sampling points,  we can show that the Jacobian matrix becomes a linear projection of a certain low-rank tensor whose rank is equal to the number of parallel branches in the model. 
This representation allows us to solve the identification problem as a tensor recovery problem, which may be approached by an alternating least squares (ALS) solution strategy.

\section{Preliminaries}

\subsection{Tensor and vector notation}
In this paper we mainly follow \cite{comon2014} in what concerns tensor notation (see also \citet{kolda2009}).
We use lowercase  ($a$) or uppercase ($A$)  plain  font for scalars,   boldface lowercase ($\vect{a}$) for vectors,  uppercase boldface ($\matr{A}$) for matrices,  calligraphic  font ($\tens{A}$) for $N$-D arrays (tensors) and script  ($\mathscr{P}$) for operators.
 Vectors are, by convention, one-column matrices. The elements of vectors/matrices/tensors are 
accessed as $a_{i} $, ${A}_{i,j}$ and $\tenselem{A}_{i_1,\ldots,i_N}$ respectively.
We use $\vecl{\cdot}$ for the standard column-major vectorization of a tensor or a matrix. 
Operator $\contr{p}$ denotes the contraction on the $p$th index of a tensor, i.e., 
\[
[\tens{A}\contr{1}\vect{u}]_{jk}=\sum_i \tenselem{A}_{ijk} u_{i}.
\]
For a matrix  ${\mA}$, ${\mA}^{\T}$ and ${\mA}^\dag$ denotes its transpose and Moore-Penrose pseudoinverse  respectively.  The notation ${\mI}_M$ is used for the $M\times M$ identity matrix and $\matr{0}_{L \times K}$ for the $L\times K$ matrix of zeroes.
We use  the symbol $\kron$ for the Kronecker product of  matrices (in order to distinguish it from the  tensor  product $\otimes$), and $\kr$ for the (column-wise) Khatri-Rao  product of matrices: i.e, the Khatri-Rao product of 
\[
\mA = \bmx \vect{a}_1 & \cdots & \vect{a}_r \emx\quad\text{and}\quad \mB = \bmx \vect{b}_1 & \cdots & \vect{b}_r \emx
\]
is defined as
\[
{\mA} \kr {\mB} =  \bmx \vect{a}_1 \kron \vect{b}_1 & \cdots & \vect{a}_r \kron \vect{b}_r
\emx.
\]
We use the notaion $\Diag{\vect{v}}$ for the diagonal matrix built from the vector $\vect{v}$.

A polyadic decomposition (PD) is a decomposition of a tensor into a sum of rank-one terms, i.e., for $\tens{Y} \in \RR^{I \times J \times K}$,
\begin{equation}\label{eq:PD}
\tens{Y} = \sum\limits_{\ell=1}^{r} \vect{a}_\ell \otimes \vect{b}_\ell \otimes \vect{c}_\ell
\end{equation}
is a polyadic decomposition. It is called canonical polyadic (CPD) if  the number  $r$  in \eqref{eq:PD} is minimal among all possible PDs of $\tens{Y}$; in that case $r$ is called the tensor rank of $\tens{Y}$.

By grouping vectors into matrices 
\[
\mA = \bmx \vect{a}_1 & \cdots& \vect{a}_r \emx, \quad \mB = \bmx \vect{b}_1 & \cdots& \vect{b}_r \emx, \quad \mC = \bmx \vect{c}_1 & \cdots& \vect{c}_r \emx
\]
we can use a more compact notation
\[
\tens{Y}=\cpd{{\mA}}{{\mB}}{{\mC}}, \quad \tenselem{Y}_{ijk} = \sum\limits_{\ell=1}^r A_{i\ell} B_{j\ell} C_{k\ell}; 
\]
for a PD (or a CPD).

Finally, for a (possibly finite) sequence 
\[
(\ldots,x(1), \ldots, x(T), \ldots)
\]
 its convolution with a vector $\vect{a} \in \mathbb{R}^{K}$  is defined as
\[
(\vect{a} \ast x )(t) = \sum\limits_{i=1}^{K} x(t-i+1) a_{i}.
\]

\subsection{Volterra kernels}
The Volterra series \citep{schetzen1980} is a classical model for nonlinear systems, and is similar to the Taylor expansion for multivariate maps.
In the discrete-time case, Volterra series can be interpreted as a power series expansion of the output of a system as a function of past inputs: $y(t) = f^{(0)} +$
\[
 \sum\limits^{\infty}_{s=1} \left( \sum^{\infty}_{\tau_1=0} \cdots\sum^{\infty}_{\tau_s=0}  H^{(s)}(\tau_1,\ldots,\tau_s) u(t-\tau_1) \cdots u(t- \tau_s)\right), 
\]
where $H^{(s)}(\cdot)$ is the $s$-th order Volterra kernel.
In the special case when the output depends only on a finite number $L$ of past inputs, i.e.,  is defined by $f: \RR^{L} \to  \RR$
\begin{equation}\label{eq:dependence_past_inputs}
y(t) = f(u({t}), u({t-1}), \ldots, u({t-L+1})),
\end{equation}
we can consider  truncated the Volterra kernels  $\tens{H}^{(s)}$ (which are $L \times \cdots \times L$ tensors). 
By denoting  for convenience the vector of past inputs as
\begin{equation}\label{eq:vect_past_inputs}
\vect{u} =  \bmx u({t}) & u({t-1}) & \ldots & u({t-L+1}) \emx^{\T}
\end{equation}
we can write the function expansion as
\begin{equation}\label{eq:expansion_truncated}
y({t}) = f(\vect{u}) = f^{(0)} + f^{(1)} (\vect{u}) + \cdots + f^{(d)} (\vect{u}) + \cdots,
\end{equation}
where the degree-$s$ term is given by
\[
f^{(s)} (\vect{u}) = \sum_{i_1,\ldots,i_s=1}^{L,\ldots, L}\tenselem{H}^{(s)}_{i_1,\ldots, i_s} u({t} - i_1+1) \cdots u({t}-i_s+1),
\]
with 
\[
\tenselem{H}^{(s)}_{i_1,\ldots, i_s} = H^{(s)}(i_1-1,\ldots,i_s-1).
\]
Order-$s$ terms can be  compactly expressed using the multiple contraction
\[
f^{(s)}(\vect{u}) = \tens{H}^{(s)} \contr{1} \vect{u} \contr{2} \vect{u}  \cdots \contr{s} \vect{u}. 
\] 

\subsection{Parallel Wiener-Hammerstein model}
In this paper, we consider the case when the LTI blocks  in Fig.~\ref{fig:pwh} are given by finite impulse response (FIR) filters of lags $L_1$ and $L_2$ respectively.
Formally, the output $y(t)$ at a time instant $t$  of a parallel Wiener-Hammerstein system is given by a composition of convolutions and univariate nonlinear transformations:
\[
y = \sum\limits_{\ell=1}^{r} \vect{b}_\ell \ast {z}_{\ell}, \quad z_{\ell} = g_{\ell}(x_\ell), \quad x_{\ell} =\vect{a}_\ell \ast u,
\]
where $\vect{a}_{\ell} \in \RR^{L_1}$, $\vect{b}_{\ell} \in \RR^{L_2}$, $g_{\ell}: \RR \to \RR$.
In this case, it is easy to see that the output $y(t)$ of the system depends only on $L = L_1 +L_2-1$ past inputs, i.e.,  $\vect{u}$ in \eqref{eq:vect_past_inputs}.

In this paper, we also add another simplifying assumption that each $g_{\ell}(t)$ is a polynomial of degree $d$. 
Therefore, the function $f$ in \eqref{eq:dependence_past_inputs} is a degree-$d$ polynomial and thus the system is completely characterized by  the first $d$ truncated  $L \times \cdots \times L$ Volterra kernels (i.e., by the collection of the homogeneous terms $f^{(s)}$ up to degree $d$, see \eqref{eq:expansion_truncated}).

\section{First-order information and projection}
\subsection{An overview of the proposed approach}
Our main idea is to exploit the first-order information in spirit of  the method in \citet{dreesen}. 
The  original method of \citet{dreesen} is designed for decoupling a static nonlinear function $f$ based on the evaluations of the first-order derivatives (Jacobians) of $f$ at a chosen set of operating points $\vect{u}^{(1)}, \ldots, \vect{u}^{(N)}$, by stacking these evaluations in a 3rd order tensor and  performing its CPD.

Note that in case of a polynomial map \eqref{eq:dependence_past_inputs}, the derivatives can be easily computed from the  Volterra kernels thanks to the following identity for degree-$s$ parts:
\begin{equation}\label{eq:volterra_gradient}
\nabla f^{(s)}(\vect{u}) = s \cdot \tens{H}^{(s)} \contr{2} \vect{u}  \cdots \contr{s} \vect{u}.
\end{equation}
However, a direct application of the decoupling technique is not possible in our case due to the following issues:
\begin{itemize}
\item the method of \cite{dreesen} does not take into account the dynamics;
\item the method of \cite{dreesen} is not applicable to single-output functions (the Jacobian tensor becomes a matrix).
\end{itemize}
Some remedies for these issues were proposed in the literature.
For example, \cite{usevich2014} reformulated the problem as structured matrix completion, \cite{hollander2017} introduced constraints on the factors of the CPD, while  \cite{lva2018} considered tensors of higher-order derivatives.
However, none of these approaches provide an out-of-the box solution for our identification problem.

In this paper, we propose an entirely different approach.
We use only the first-order information of $f$; however, we split $f$ into homogeneous parts \eqref{eq:expansion_truncated} in the spirit of \citet{vanmulders2014}.
A particular choice of tailored operating points (see subsection~\ref{sec:tailored_points}) allows us to show that the vector of evaluations of the gradients of the homogeneous parts can be viewed as a linear projection (sampling) of a third-order tensor $\tens{T}$ whose rank is equal to the number of branches and whose factors give the coefficients for the filters in the LTI blocks.
This allows us to reformulate the identification problem as a low-rank tensor recovery problem.

The remainder of this paper is organized as follows.
In the current section, we focus only on the case of a single branch.
We begin by some preliminary observations, followed by  describing the structure of the tailored operating points.
For such points, we then describe the building blocks for the projection operator in subsection~\ref{sec:grad_proj} and show in subsection~\ref{sec:proj_rankone} that the vector of gradient evaluations is a projection of a rank-one tensor.
The overall algorithm for $r$ branches is presented in section~\ref{sec:tensor_recovery}, where an algorithm for tensor recovery is also dicussed.
The numerical experiments are provided in section~\ref{sec:experiments}.

\subsection{Single branch: preliminary observations}
We consider the case of a single branch, with the filters  
\[
\vect{a} = \bmx a_{1}, \ldots, a_{L_1}\emx^{\T},\quad \vect{b} = \bmx b_{1}, \ldots, b_{L_2}\emx^{\T}
\]
and the single (not necessarily homogeneous) nonlinearity  $g(t)$.
Then,  the output of a single branch is given by \eqref{eq:dependence_past_inputs} with the nonlinear map
\[
f(\vect{u}) = \vect{b}^{\top} \vect{g}(\matr{V}^{\T} \vect{u}),
\]
where $\vect{g}(\cdot)$ is defined as
\[
\vect{g}(x_1,\ldots,x_{L_1}) = \bmx g(x_1) &  \ldots & g(x_{L_1})\emx^{\T},
\] 
and  $\matr{V} \in \mathbb{R}^{L\times L_2}$ is the following Toeplitz matrix:
\begin{equation}\label{eq:V_matrix}
\matr{V}  =
\bmx
\vect{v}_1 & \cdots & \vect{v}_{L_2}
\emx = 
\bmx
a_1     &  &  \\
\vdots  & \ddots  &  \\
a_{L_1}     &  & a_{1}  \\
             & \ddots  & \vdots  \\
           &   & a_{L_1} \\
\emx.
\end{equation}
By the chain rule (as in \citet{dreesen}), the gradient of  $f$ has the form 
\begin{equation}\label{eq:gradient_single_branch}
\nabla f (\vect{u}) = \matr{V} \bmx g'(\matr{v}_1^{\T} \vect{u}) & & \\ & \ddots & \\&&g'(\matr{v}_{L_2}^{\T} \vect{u})  \emx \vect{b}.
\end{equation}
\begin{rem}
Although the function $f$, the model of the dynamical system, and the Volterra kernels were initially defined for  real inputs, the expressions in \eqref{eq:volterra_gradient} and \eqref{eq:gradient_single_branch} are polynomial in $\vect{u}$, hence we can formally evaluate them at complex points $\vect{u} \in \mathbb{C}^{L}$.
This is one of the important features of our approach that allows us to avoid some numerical issues.
\end{rem}

\subsection{Tailored operation points}\label{sec:tailored_points}
Next, we restrict our attention to homogeneous parts of the nonlinearity $g(x) = cx^s$.
Another key idea of our method is to use tailored operating points $\vect{u} \in\mathbb{C}^{L}$ in order to simplify the expression in \eqref{eq:gradient_single_branch}.
We are going to use Vandermonde-like operating points parameterized by $\mu \in \mathbb{C}$:
\[
\vect{u}_{\pt} =  \bmx {1} & \pt & \pt^2 & \ldots & \pt^{L-1}\emx^{\T}.
\]
In this case, it is not difficult to see that 
\begin{align*}
g'(\matr{v}_k^{\T} \vect{u}_{\pt}) &= c s (\matr{v}_k^{\T} \vect{u}_{\pt})^{s-1} \\
&= c s (\pt^{k-1} a(\pt))^{s-1} = c s \pt^{(k-1)(s-1)} (a(\pt))^{s-1},
\end{align*}
where $a(\pt) = a_1 + a_2 \pt + \cdots + a_{L_1} \pt^{L_1-1}$. Plugging this expression in \eqref{eq:gradient_single_branch}, we  obtain that
\begin{equation}\label{eq:gradient_single_homogeneous}
\nabla f^{(s)} (\vect{u}_{\pt}) = (c s (a(\pt))^{s-1}) \vect{v}_{\pt}
\end{equation}
where the vector  $\vect{v}_{\pt} \in \mathbb{C}^{L}$ is defined as
\begin{align*}
&\vect{v}_{\pt} = \matr{V} 
\bmx b_{1} \\ b_{2} \pt^{(s-1)} \\ \vdots \\ b_{L_2}\pt^{(L_2-1)(s-1)}\emx\\
&= 
\bmx
a_1 b_1 \\
a_2 b_1 + a_1 b_2 \pt^{(s-1)} \\
a_3 b_1 + a_2 b_2 \pt^{(s-1)} +  a_1 b_3 \pt^{2(s-1)}\\
\vdots \\
 a_{L_1} b_{L_2-1} \pt^{(L_2-2)(s-1)} + a_{L_1-1} b_{L_2} \pt^{(L_2-1)(s-1)}\\
 a_{L_1} b_{L_2} \pt^{(L_2-1)(s-1)}
\emx,
\end{align*}
where $\matr{V}$ is defined in \eqref{eq:V_matrix}.

\subsection{Gradient as a projection of a rank-one term}\label{sec:grad_proj}
We are going to show that $\vect{v}_{\pt}$ from the previous subsection can be conveniently written as a linear projection (sampling) of a rank-one matrix. First of all, we introduce the diagonal summation (``Hankelization'') operator
$\mathscr{H}: \CC^{L_1 \times L_2} \to \CC^{L}$, which takes the sums on the antidiagonals
\[
\mathscr{H} (\matr{E}) = 
\bmx
E_{1,1} \\
E_{2,1} + E_{1,2} \\
E_{3,1} + E_{2,2} +  E_{1,3}\\
\vdots \\
 E_{L_1,L_2-1}  + E_{L_1-1,L_2}\\
 E_{L_1,L_2}
\emx.
\]
Next, it is easy to see that $\vect{v}_{\pt}$ can be obtained by applying the projection operator $\ProjHD{\pt}{s}: \CC^{L_1 \times L_2} \to \CC^{L}$, which is a composition of the diagonal summation with  the scaling of columns by powers of $\pt$:
\[
\ProjHD{\pt}{s}(\matr{E})  = \mathscr{H} (\matr{E} \Diag{\bmx 1 & \pt^{(s-1)} & \cdots & \pt^{(L_2-1)(s-1)}\emx}),
\]
i.e., $\vect{v}_{\pt} =\ProjHD{\pt}{s}( \vect{a}\vect{b}^{\T})$.
After that, we get that the gradient in \eqref{eq:gradient_single_homogeneous} can be expressed as follows
\[
\nabla f^{(s)} (\vect{u}_{\pt}) = \ProjHD{\pt}{s}((c s (a(\pt))^{s-1}) \vect{a}\vect{b}^{\T} ) .
\]
Finally, in the next subsections we are going  to evaluate the gradients at different operating points and collect  information from several kernels. 
\subsection{Combining several kernels and points}\label{sec:proj_rankone}
Now consider a set of $N$ points in the complex plane
\[
\{\pt_1, \ldots, \pt_N\} \subset \CC
\]
at which we are evaluating the gradients of the homogeneous parts, and collecting them into one single vector:
\begin{equation}\label{eq:gradient_vector}
\vect{y} = 
 \bmx
\nabla f^{(1)} (\vect{u}_{1}) \\
\nabla f^{(2)} (\vect{u}_{\pt_1}) \\
\vdots \\
\nabla f^{(2)} (\vect{u}_{\pt_N}) \\
\vdots \\
\nabla f^{(d)} (\vect{u}_{\pt_1}) \\
\vdots \\
\nabla f^{(d)} (\vect{u}_{\pt_N}) \\
\emx \in \CC^{((d-1)N+1)L}.
\end{equation}
Unlike the previous section, we now consider  a general polynomial nonlinearity:
\[
g(t) = c_1 t + c_2 t^2 + \cdots + c_d t^d.
\]
By using the results of the previous subsection, we can show that $y$ is a projection of a rank-1 tensor:
\[
\vect{y} = \mathscr{P}(\tens{T}),
\]
where the rank-one tensor is
\[
\tens{T} = \vect{a} \otimes \vect{b} \otimes \vect{h},
\]
vectors $\vect{a}$, $\vect{b}$ are as before,  and $\vect{h}^{\T} = $
\[
\bmx 
c_1 &
2c_2  a(\pt_1) &\!\!
\cdot\cdot &
2 c_2 a(\pt_N) &\!\!
\cdot\cdot &
dc_d  (a(\pt_1))^{d\text{-}1} &
\!\!\cdot\cdot &
d c_d (a(\pt_N))^{d\text{-}1} 
\emx.
\]
The sampling operator is defined as a concatenation of sampling operators of tensor slices
\[
\mathscr{P}(\tens{T}) = 
 \bmx
\ProjHD{1}{1} (\tens{T}_{:,:,1}) \\
\ProjHD{\pt_1}{2} (\tens{T}_{:,:,2}) \\
\vdots \\
\ProjHD{\pt_N}{2}  (\tens{T}_{:,:,1+N}) \\
\vdots \\
\ProjHD{\pt_1}{d}  (\tens{T}_{:,:,2+N(d-2)}) \\
\vdots \\
\ProjHD{\pt_N}{d}  (\tens{T}_{:,:,1+N(d-1)}) \\
\emx.
\]

\section{Identification as tensor recovery}\label{sec:tensor_recovery}

\subsection{Several branches and overall algorithm}
We saw in the previous section that in the case of a single branch, the vector of the gradients $\vect{y}$ evaluated at the Vandermonde evaluation points $\vect{u}_{\pt_k}$, $k = 1,\ldots, N$, is a projection of a rank-one tensor.
This implies that for a sum of $r$ branches the vector $\vect{y}$ is a projection of a tensor having polyadic decomposition with $r$ terms:
\[
\vect{y} = \mathscr{P}(\tens{T}), \quad \tens{T} = \cpd{\mA}{\mB}{\mH}=\sum\limits_{\ell=1}^{r} \vect{a}_\ell \otimes \vect{b}_\ell \otimes \vect{h}_\ell,
\]
where $\vect{a}_\ell$, $\vect{b}_\ell$ are the coefficients of the corresponding filters, and $\vect{h}_\ell$ are the vectors for the nonlinearities constructed as previously. 
Thus the identification problem can be reformulated as a low-rank tensor recovery of the tensor $\tens{T}$ from the samples $\vect{y}$.
Low-rank tensor recovery is a generalization of the tensor completion problem to the case of arbitrary  sampling operators (and not just selection of the elements as in a typical tensor completion problem).
%The tensor recovery problem is similar in spirit to  matrix sensing problems.

This leads us to the following algorithm.
\begin{alg}\label{alg:pwh_volterra_sampling}
Input: number of branches $r$, filter sizes $L_1,L_2$, Volterra kernels up to order $d$.
\begin{enumerate}
\item Choose sampling points $\pt_1,\ldots, \pt_N \in \CC$.
\item Evaluate the gradients of the homogeneous parts of $f$ at $\vect{u}_{\pt_k}$ via contractions with the Volterra kernels (see \eqref{eq:volterra_gradient}).
\item Build $\vect{y}$ as in (\ref{eq:gradient_vector}) by evaluating the gradients via Volterra kernels.
\item Find the rank-$r$ tensor $\tens{T} = \cpd{\mA}{\mB}{\mH}$ such that $\vect{y} \approx \mathscr{P}(\tens{T})$.
\item Recover the filter coefficients from $\vect{a}_{\ell}$, $\vect{b}_{\ell}$.
\item Recover the coefficients of the polynomials 
\[
g_\ell(t) = c_{\ell,1} t + c_{\ell,2} t^2 + \cdots + c_{\ell,d} t^d
\]
by solving
\[
\vect{h}_\ell \approx \bmx 
c_{\ell,1} \\
2c_{\ell,2}  (a_\ell(\pt_1)) \\
\vdots \\
2 c_{\ell,2} (a_\ell(\pt_N)) \\
\vdots \\
dc_{\ell,d}  (a_\ell(\pt_1))^{d-1} \\
\vdots \\
d c_{\ell,d} (a_\ell(\pt_N))^{d-1} 
\emx
\]
\end{enumerate}
\end{alg}

\begin{rem}
In order to avoid numerical issues we  restrict the sampling points to the unit circle 
\[
\mathbb{T} = \{\pt\in \mathbb{C} : |\pt| = 1\}.
\]
\end{rem}
Also, in Algorithm~\ref{alg:pwh_volterra_sampling}, we allow for approximations of $\vect{y}$ in order to account for modelling errors or noise.
While the estimation of $g_{\ell}(t)$ is a simple least squares problem, the most difficult part becomes the CPD of a partially observed tensor, which we detail in the next section.

\subsection{Partially observed CPD}
In order to find the rank-$r$ tensor from its projection, we are going to solve the following tensor recovery problem in the least squares sense:
\[
\min_{{\mA},{\mB},{\mH}} \|\mathscr{P} (\cpd{\mA}{\mB}{\mH}) - \vect{y} \|_2^2,
\]
where $\mathscr{P}: \CC^{L_1 \times L_2 \times L_3} \to \CC^{M}$ is a sampling operator.

We are going to use a well-known alternating least squares (block coordinate descent) strategy \cite{kolda2009}. This strategy consists in alternate minimization with respect to each variable with fixing  other variables, and can be summarized in the following algorithm.
\begin{alg}[Partial ALS]\label{alg:als}
Input: initializations  $\mA_0$, $\mB_0$, $\mH_0$.
\begin{enumerate}
\item For k=1,2,....  until a stopping criterion is satisfied
\item \quad $\mA_{k} \leftarrow \arg\min_{{\mA}}\|\mathscr{P} (\cpd{\mA}{\mB_{k-1}}{\mH_{k-1}}) - \vect{y} \|_2^2$;
\item \quad $\mB_{k} \leftarrow \arg\min_{{\mA}}\|\mathscr{P} (\cpd{\mA_k}{\mB}{\mH_{k-1}}) - \vect{y} \|_2^2$;
\item \quad $\mH_{k} \leftarrow \arg\min_{{\mA}}\|\mathscr{P} (\cpd{\mA_k}{\mB_k}{\mH}) - \vect{y} \|_2^2$.
\item End for 
\end{enumerate}
\end{alg}
Each update in Algorithm~\ref{alg:als} is a  linear least squares problem, which explains the name ``alternating least squares''.
Note that the overall cost function is nonconvex, and thus the algorithm may suffer from local minima and other convergence problems \citep{comon2009tensor}.
However, this is one of the most popular and practically successful strategies.
In what follows, we provide details on   implementation of updates for recovery of partially observed low-rank tensors, which we did not find in the literature. 

We assume that the operator $\mathscr{P}: \CC^{L_1 \times L_2 \times L_3} \to \CC^{M}$ has the  matrix representation $\matr{P} \in \CC^{M \times (L_1  L_2  L_3)}$, i.e.,
 \[
 \mathscr{P}(\tens{T}) = \matr{P}  \vecl{\tens{T}}.
 \]
 Then the updates of ALS can be derived as follows:
 \begin{itemize}
 \item Updating $\matr{A}$: \quad $\vecl{\mA} =  (\mZ_{\mA})^{\dagger} \vect{y}$, where
 \[
 \mZ_{\mA} =  \mP ((\mC \kr \mB) \kron \mI_{L_1}).
 \]
 \item Updating $\matr{B}$:\quad $\vecl{\mB} =  (\mZ_{\mB})^{\dagger} \vect{y}$, where
 \[
\mZ_{\mB} = \mP \bmx \vect{c}_1 \kron \mI_{L_2} \kron \vect{a}_1 & \cdots & \vect{c}_r \kron \mI_{L_2} \kron \vect{a}_r  \emx.
 \]
 
 \item Updating $\matr{C}$:\quad $\vecl{\mC} =  (\mZ_{\mC})^{\dagger} \vect{y}$, where
 \[
 \mZ_{\mC} = \mP \bmx \mI_{L_3} \kron \vect{b}_1 \kron \vect{a}_1 & \cdots & \mI_{L_3} \kron \vect{b}_r \kron \vect{a}_r  \emx. 
 \]
 \end{itemize}
For the practical implementation, we take advantage of the sparsity of the matrix $\matr{P}$: an easy inspection reveals that $\matr{P}$ is block-diagonal with  banded blocks.

\section{Experiments}\label{sec:experiments}
Here we present an example that illustrates our approach. 
The algorithms were implemented in MATLAB  R2019b on MacBook Air  (2014, 1.4 GHz Intel i5, 4GB RAM).

We consider $r = 2$ branches and filter lengths $L_1 = 3$, $L_2 = 3$ with the following coefficients:
\[
\mA =  
\bmx
0.3 & 0.6 \\
-0.4 & 0.2 \\
0.1 & 0.3
\emx,
\quad
\mB = 
\bmx
0.3 & 0.2 \\
0.2 & 0.3 \\
0.1 & 0.01
\emx,
\]
and nonlinearities
\begin{equation}\label{eq:example_poly}
g_1(x_1) = 3x_1^3-x_1^2+5, \quad  g_2(x_2) = -5x_2^3+3x_2-7.
\end{equation}
\begin{figure}
\begin{center}
\includegraphics[width=0.5\textwidth]{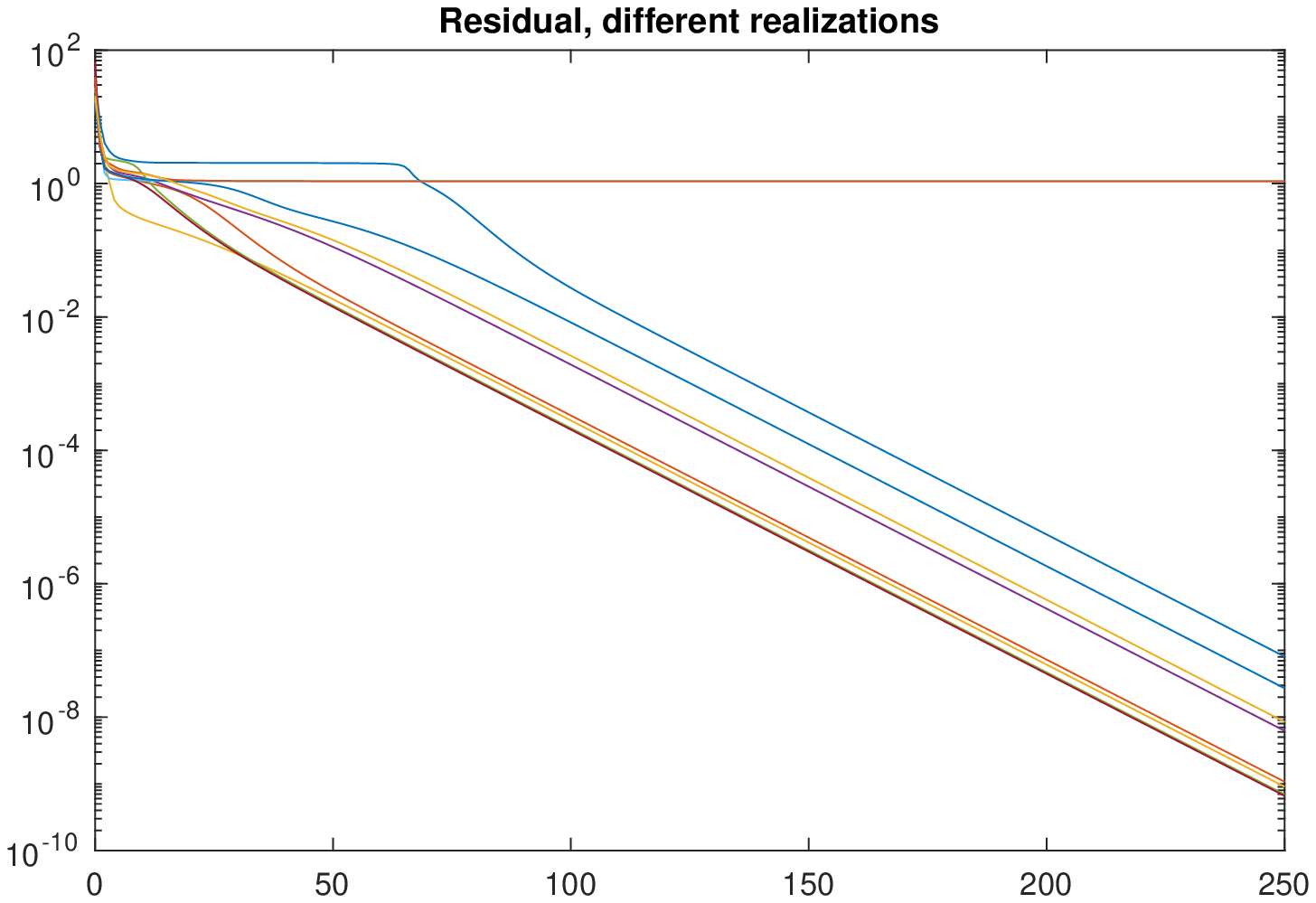}
\caption{Evolution of the residual ($\|\mathscr{P} (\cpd{\mA}{\mB}{\mH}) - \vect{y} \|_2$) with respect to the number of cycles of ALS.} 
\label{fig:cost}
\end{center}
\end{figure}%
We use $N=30$ operating points generated randomly on the unit circle. 
We run  Algorithm~\ref{alg:pwh_volterra_sampling} for $10$ different starting points (i.i.d. Gaussian distributed), maximum $250$ iterations, and show the convergence plots in Fig.~\ref{fig:cost}.
We see that the algorithm converges linearly for all but one initialization, which is reasonable due to nonconvexity of the problem.
For one of the realizations, the final residual is $\|\mathscr{P} (\cpd{\mA}{\mB}{\mH}) - \vect{y} \|_2= 8.48 \cdot 10^{-9}$, and the estimated factor $\widehat{\matr{A}}$ is (with the first row normalized to $1$ and shown with $4$ fractional digits of the mantissa),
\[
\widehat{\matr{A}} = 
\bmx
1 &1 \\
 -1.3333 - i0.8305 \cdot 10^{-8} &  0.3333 + i0.4203\cdot 10^{-9}\\
  0.3333 - i0.2210 \cdot 10^{-8}&  0.4999 + i0.7007 \cdot10^{-10}\\
\emx,
\]
which is complex-valued, but recovers quite accurately the true ${\matr{A}}$ (the same holds for $\matr{B}$, not shown here).

In order to illustrate the reconstruction of the nonlinearities, instead of solving the least squares problem in Algorithm~\ref{alg:pwh_volterra_sampling}, we apply the idea similar the visualization of nonlinearities in \citet{dreesen}. In fact, the elements of $\mH$ can be combined in such a way to yield the values of the derivatives of $g_{\ell}(\cdots)$ at the points $a(\pt_k)$.
We perform polynomial regression for degree $2$, take the real parts and obtain the following polynomials (with leading coefficient normalized to $1$), rounded to the $4$ fractional digits
\[
h_1(t) = t^2 - 0.2222 t, \quad h_2(t) =  t^2 - 0.2.
\]
after inspecting \eqref{eq:example_poly}, we obtain that these are  (up to numerical errors) the derivatives of the original nonlinearities, (i.e., $h_1(t) = \alpha g_{1}'(t)$, $h_2(t) = \alpha g_{2}'(t)$ with $\alpha = 9$).

\section{Conclusion}
We developed a novel promising algorithm for identification of Wiener-Hammerstein systems from Volterra kernels.
Our approach has the  following advantages:
\begin{itemize}
\item It is based on tensor recovery, rather than CPD with structured factors, and can be solved with a simple alternating least squares scheme.
\item It does not need all the coefficients of the Volterra kernels to be estimated: we just need to compute contractions with Vandermonde-structured vectors for a fixed number of operating points.
\end{itemize}
Furthermore, we believe that our method may have an interesting interpretation from the frequency-domain identification perspective.
Note that the operating points that we use are typically taken on the unit circle, i.e., an operating point is chosen as $\pt_k = e^{2\pi i\omega_k}$.
Viewed from a frequency-domain point of view \citep{pintelon2012},  contraction of  Volterra kernels with Vandermonde-structured vectors is somewhat similar to an ``excitation'' of the first-order derivative at a frequency $\omega_k$.
However, for such an interpretation, we would potentially need to consider the framework of the Volterra kernel identification with complex valued inputs \citep{bouvier2019ijc}.

\begin{ack}
This  research  was  supported  by the  ANR (Agence Nationale de Recherche) grant LeaFleT (ANR-19-CE23-0021); KU Leuven start-up-grant STG/19/036 ZKD7924; KU Leuven Research Fund; Fonds Wetenschappelijk Onderzoek - Vlaanderen (EOS
Project 30468160 (SeLMA), SBO project S005319N, Infrastructure project
I013218N, TBM Project T001919N, Research projects G028015N, G090117N,
PhD grants SB/1SA1319N, SB/1S93918, and SB/151622); Flemish Government (AI
Research Program); European Research Council under the European
Union’s Horizon 2020 research and innovation programme (ERC AdG grant
885682). P. Dreesen is affiliated to Leuven.AI -- KU Leuven institute for AI,
Leuven, Belgium.
Part of this work was performed while P. Dreesen and M. Ishteva were with Dept. ELEC of Vrije Universiteit Brussel, and P. Dreesen was with CoSys-lab at Universiteit Antwerpen, Belgium.
The authors would like to thank the three anonymous reviewers for their useful comments that helped to improve the presentation of the results.
\end{ack}

\bibliography{sysid-volterra}             % bib file to produce the bibliography
                                                     % with bibtex (preferred)
                                                   
%\begin{thebibliography}{xx}  % you can also add the bibliography by hand

%\bibitem[Able(1956)]{Abl:56}
%B.C. Able.
%\newblock Nucleic acid content of microscope.
%\newblock \emph{Nature}, 135:\penalty0 7--9, 1956.

%\bibitem[Able et~al.(1954)Able, Tagg, and Rush]{AbTaRu:54}
%B.C. Able, R.A. Tagg, and M.~Rush.
%\newblock Enzyme-catalyzed cellular transanimations.
%\newblock In A.F. Round, editor, \emph{Advances in Enzymology}, volume~2, pages
%  125--247. Academic Press, New York, 3rd edition, 1954.

%\bibitem[Keohane(1958)]{Keo:58}
%R.~Keohane.
%\newblock \emph{Power and Interdependence: World Politics in Transitions}.
%\newblock Little, Brown \& Co., Boston, 1958.

%\bibitem[Powers(1985)]{Pow:85}
%T.~Powers.
%\newblock Is there a way out?
%\newblock \emph{Harpers}, pages 35--47, June 1985.

%\bibitem[Soukhanov(1992)]{Heritage:92}
%A.~H. Soukhanov, editor.
%\newblock \emph{{The American Heritage. Dictionary of the American Language}}.
%\newblock Houghton Mifflin Company, 1992.

%\end{thebibliography}

%\appendix
%\section{A summary of Latin grammar}    % Each appendix must have a short title.
%\section{Some Latin vocabulary}              % Sections and subsections are supported  
                                                                         % in the appendices.
\end{document}